\documentclass[12pt, review, authoryear]{article}

\makeatletter
\def\ps@pprintTitle{%
 \let\@oddhead\@empty
 \let\@evenhead\@empty
 \def\@oddfoot{\centerline{\thepage}}%
 \let\@evenfoot\@oddfoot}
\makeatother

\usepackage[english]{babel}
\usepackage[utf8x]{inputenc}

\usepackage[a4paper,top=2cm,bottom=2cm,left=1.45cm,right=2cm,marginparwidth=1.5cm]{geometry} 
\usepackage{float}
\usepackage{relsize}
\usepackage{lipsum}
\usepackage{setspace}
\usepackage{multicol}
\usepackage{lineno}
\usepackage{amsmath}
\usepackage{amssymb}
\usepackage{bm}
\usepackage{bbm}
\usepackage{verbatim}
\usepackage[ruled]{algorithm2e}
\usepackage{graphicx}
\usepackage{rotating}
\usepackage{outlines}
\usepackage{authblk}
\usepackage[round]{natbib}
\usepackage[colorinlistoftodos, disable]{todonotes}   %disable
\usepackage[colorlinks=true, allcolors=blue]{hyperref}
\usepackage{cleveref}   %always the last one

\usepackage{soul}
\usepackage{ulem} % for strikethrough: \sout{text}
\graphicspath{{Figures/Thesis_Figures/}}

\newcommand\todoDONE[1]{\todo[backgroundcolor=green]{#1}}

\DeclareMathOperator{\Cov}{\mathrm{Cov}}

\usepackage{etoolbox}
\patchcmd{\pprintMaketitle}%
{\vskip36pt} % original vertical space 36 pt
{\vskip10pt} % minimum vertical space  <<<<<<<<<<<
{}
{}
\newsavebox\extrainfobox

\makeatletter
\renewenvironment{abstract}
 {\small
  \begin{center}
  \bfseries \abstractname\vspace{-0.5em}\vspace{0pt}
  \end{center}
  \list{}{
    \setlength{\leftmargin}{0cm}%
    \setlength{\rightmargin}{\leftmargin}%
  }%
  \item\relax}
 {\endlist}

\title{Normative brain mapping of 3-dimensional morphometry imaging data using skewed functional data analysis}

\author[1]{Marco Palma\thanks{Corresponding author: marco.palma@mrc-bsu.cam.ac.uk}}
\author[2]{Shahin Tavakoli}
\author[3,4]{Julia Brettschneider}
\author[5]{Ana-Maria Staicu}
\author[6,7]{Thomas E. Nichols}
\author[ ]{for the Alzheimer's Disease Neuroimaging Initiative\footnote{Data used in this work were obtained from the Alzheimer's Disease Neuroimaging Initiative (ADNI) database (\url{http://adni.loni.usc.edu}). As such, the investigators within the ADNI contributed to the design and implementation of ADNI and/or provided data but did not participate in analysis or writing of this work. A complete listing of ADNI investigators can be found at: \url{http://adni.loni.usc.edu/wp-content/uploads/how_to_apply/ADNI_Acknowledgement_List.pdf}.}}

\affil[1]{MRC Biostatistics Unit, University of Cambridge, Cambridge, CB2 0SR, United Kingdom}
\affil[2]{Research Center for Statistics, Geneva School of Economics and Management, University of Geneva, 1205 Geneva, Switzerland}
\affil[3]{Department of Statistics, University of Warwick, Coventry, CV4 7AL, United Kingdom}
\affil[4]{The Alan Turing Institute, London, NW1 2DB, United Kingdom}
\affil[5]{Department of Statistics, North Carolina State University, Raleigh, North Carolina 27695, United States}
\affil[6]{Oxford Big Data Institute, Li Ka Shing Centre for Health Information and Discovery, Nuffield Department of Population Health, University of Oxford, Oxford, OX3 7LF, United Kingdom}
\affil[7]{Wellcome Centre for Integrative Neuroimaging, FMRIB, Nuffield Department of Clinical Neurosciences, University of Oxford, Oxford, OX3 9DU, United Kingdom}

\begin{document}

\maketitle

\begin{abstract}

Tensor-based morphometry (TBM) aims at showing local differences in brain volumes with respect to a common template. TBM images are smooth but they exhibit (especially in diseased groups) higher values in some brain regions called lateral ventricles. More specifically, our voxelwise analysis shows both a mean-variance relationship in these areas and evidence of spatially dependent skewness. %which can be missed in the standard functional data analysis (FDA) settings, which are focused only on the first two functional moments through functional principal component analysis. 

We propose a model for 3-dimensional functional data where mean, variance, and skewness functions vary smoothly across brain locations. We model the voxelwise distributions as skew-normal. The smooth effects of age and sex are estimated on a reference population of cognitively normal subjects from the Alzheimer's Disease Neuroimaging Initiative (ADNI) dataset and mapped across the whole brain. 

The three parameter functions allow to transform each TBM image (in the reference population as well as in a test set) into a Gaussian process. These subject-specific normative maps are used to derive indices of deviation from a healthy condition to assess the individual risk of pathological degeneration.

\end{abstract}

\doublespacing

\newpage

\section{Introduction}

\begin{comment}
\begin{outline}
    \1 Studying shapes and volumes of the brain
    \1 Intro to TBM
    \1 TBM in ADNI: motivation
    \1 TBM and voxelwise distributions: literature review
    \1 Our approach
        \2 Provide an atlas of the mean TBM images in ADNI for cognitive normal subpopulation
        \2 Derive sex- and age-related effects 
        \2 Introduce z-maps referenced to the mean of normal population
        \2 Derive subject-specific indices of “abnormality” 
\end{outline}

\begin{outline}
    \1 Studying shapes and volumes of the brain
    \1 Intro to TBM
    \1 Studying TBM and in general brain images in the healthy group: normative modelling \citep{bethlehem2022brain, marquand2016beyond}
    \1 Growth charts \citep{bethlehem2022brain,chen2015quantile}
    \1 Our approach
\end{outline}

\end{comment}

The study of shapes and volumes of brain regions represents a valid approach to highlight differences between subjects \citep{ashburner2004morphometry}. Many phenomena,  non-pathological (like ageing) as well as pathological (e.g.\ Alzheimer's disease), are characterised by increasing atrophy at differential rates throughout the brain lobes. The atrophy is shown (cross-sectionally between subjects or longitudinally) through deformation of structural magnetic resonance images (sMRI).

Within the family of brain morphometry methods, tensor-based morphometry (TBM) is used to identify regional volumetric deviations from a common sMRI template \citep{hua2013unbiased}. %Each voxel in TBM images is associated to the relative volumetric differences with respect to an  template. 
The numeric value at each brain location can be interpreted as a multiplicative factor of expansion or shrinkage of the brain area. In particular, values above 1 in a brain area indicate that the subject shows an expanded volume with respect to the common template: for example, a TBM value of 1.1 means that the volume in the voxel of the subject image is $10\%$ higher than the volume in the same voxel in the common template. %This multiplicative factor of expansion/contraction corresponds to the determinant of the Jacobian matrix that for each voxel encodes the deformation that maps the points in the template to the original MRI scan of the subject \citep{ashburner2004morphometry, chung2013computational}. 
For each voxel, this multiplicative factor of expansion/contraction is computed as the determinant of a Jacobian matrix which represents the alignment between the MRI image and the template \citep{ashburner2004morphometry, chung2013computational}.
To model the distribution of TBM values, two options are described in the literature \citep{chung2013computational}: the normal distribution \citep{chung2003deformation}, and the log-normal distribution (several arguments in favour of this option, including a better account of skewness, are listed in \citealp{leow2007statistical}). Nevertheless, the features of the voxelwise distributions have not been largely explored in the literature about TBM imaging. In particular, it is unclear also how those features vary between subjects in the healthy population and patients. 

%not straightforward. In \citet{chung2013computational} two alternatives are illustrated: normal distributions were traditionally assumed \citep{chung2003deformation}, while \citet{leow2007statistical} among others have discussed mathematical arguments in favour of the log-normal distribution, which also better accounts for the skewness observed in real data.

Normative modelling is a framework proposed in psychiatric applications for studying population heterogeneity \citep{rutherford2022normative} while
%the normative modelling framework \citep{rutherford2022normative} was proposed to study the heterogeneity within a neuroimaging cohort. It can be used to produce 
returning individual predictions based on a reference population \citep{marquand2016understanding, marquand2019conceptualizing}. Normative modelling is a suitable approach when there is no clear-cut separation between the groups of healthy subjects and patients. For example, different subjects might show some aspects of the disease which could require the definition of subgroups of the disease or even a broader continuous spectrum of the pathology. In some diseases, it could also be argued that the disease cluster cannot be clearly separated from the healthy subpopulation \citep{marquand2016beyond}. These views about disease as a condition with large heterogeneity are not captured in the usual case-control approach, which is useful for comparing the averages in the two clinical groups, but does not focus on the individual variation (which is often seen as a ``residual'' under that framework).
%Normative models using the healthy population as reference can be used to quantify heterogeneity in a disease-free setting. Then, subjects with no previous diagnosis could obtain a risk score that could be used to perform inferences about the individual diagnostic labels, where the disease would appear as ``extreme'' with respect to the healthy group. This might be especially useful in AD studies, where the neurodegeneration could be seen as a continuous process rather than a step function with 2 or 3 levels.

When age is included in the model, the normative approach can also be used to build for neuroimaging outcomes the analogous of the growth charts used in healthcare settings. This approach was presented in \citet{bethlehem2022brain}, where the evolution of seven brain phenotypes (dealing with tissue volumes and cortical summaries) across the human lifespan was modelled on a large aggregated sample of MRI images from multiple data sources. The brain charts were used to derive milestones in the healthy brain development but also to evaluate how ``extreme'' the phenotypes of diseases groups appeared with respect to the median of the normative population. A percentile score was used to quantify those differences, in a similar fashion as for the quantile rank maps proposed in \citet{chen2015quantile} in a study of functional connectivity.

In this work, we present a strategy for normative modelling of 3-dimensional brain imaging data showing asymmetry in the voxelwise distributions (in other words, the distributions across subjects of TBM values observed for each voxel). Following the functional data analysis (FDA; \citealp{RamSil2005}) approach illustrated in \citet{staicu2012skewedFDA} and extended in \citet{li2015csfm}, the problem is split into two components: marginal distributions to model the voxelwise features and a copula to model the dependence structure between voxels. We focus on the voxelwise distributions, which are modelled using skew-normal distributions \citep{azzalini2013skew}, that offer a flexible generalisation of the normal distribution by means of a single skewness parameter. We also specify the mean parameter as a function of age and sex and use a convenient basis function specification.

%We identify sex differences as well as age-related volume changes in the brain which are not related to pathological neurodegeneration. The functional parameters 

%Our modelling strategy allow to fully display the heterogeneity of regional brain volume changes within the healthy reference population, by means of a sex-specific, age-adjusted mean as well as the other parameters of the voxelwise distributions. We are also able to derive the expected TBM image for a healthy individual with a given age and sex (top box of \Cref{fig:NormMap_flowchart}). 
%this modelling strategy returns a mapping which could be broadly interpreted as a 3D equivalent of the growth charts which are commonly used to measure healthy ranges for height and weight in children\todo{cite}. The idea of building growth charts for brain data has been recently explored by \citet{bethlehem2022brain}, that proposed a normative model for grey matter, white matter total volumes as well as other brain phenotypes over the whole lifespan by aggregating a large range of MRI datasets. Just as in growth charts, with the functional parameters estimated in the models we identify for each voxel the range of typical values in the healthy population. 

\begin{figure}[hbtp]
    \centering
    \includegraphics[width = 0.99\linewidth]{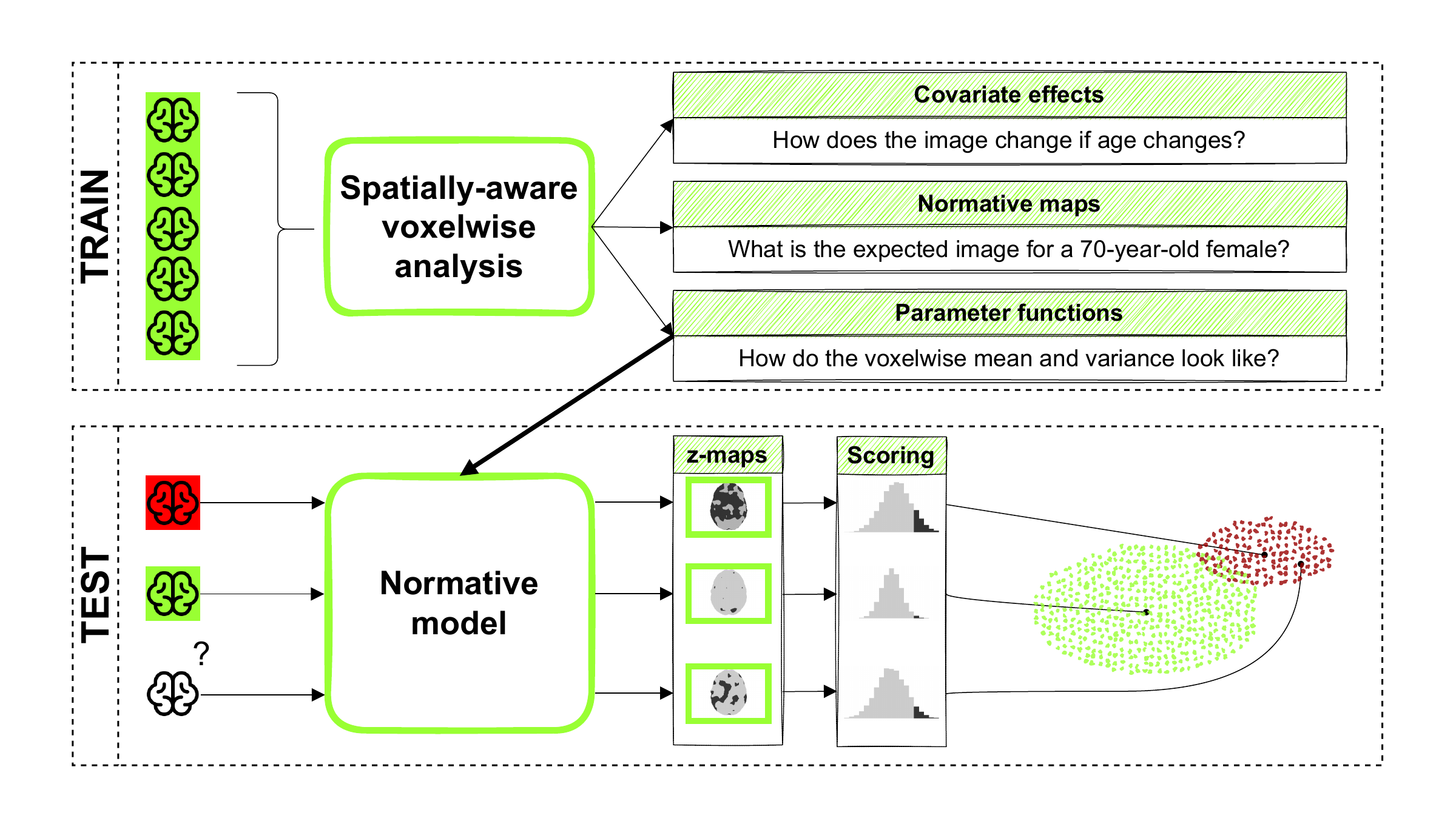}
    \caption{Scheme of the analysis. In the training phase, the parameters of the normative distribution are estimated for each voxel in a set of TBM images from healthy subjects (in green) and smoothed across the brain (spatially-aware voxelwise analysis). The voxelwise analysis allows to derive sex-specific, age-adjusted mean and to predict the expected image for specific values of the covariates. In the test phase, the normative model (that is fully defined by the parameter functions in the training phase) is used to transform the brain images of subjects with a disease (in red) or yet undiagnosed (identified with question mark) into z-maps which are referenced to the healthy population. For each subject, the values in the z-maps are plotted in a histogram and a score is defined (based for example on the values greater than a given threshold) to quantify the ``extremity'' with respect to the mean of the normative population.}
    \label{fig:NormMap_flowchart}
\end{figure}

In the context of this work, the goal of the 3D normative mapping is twofold. First, we want to describe the heterogeneity of regional brain volume changes within the healthy reference population (top panel of \Cref{fig:NormMap_flowchart}). Second, we aim at providing a model to assess how far the TBM image of a subject (not included in the training set) is from the reference population (bottom panel of \Cref{fig:NormMap_flowchart}). For the latter purpose, we transform the original TBM images into ``z-maps'' which are based on normal distributions. In this way, we expect subjects with clinically measurable neurodegeneration to appear as extremes with respect to the normative population. The information of a z-map is then summarised into a single value, capturing the extent of the departure of a subject from the normative population in the form of a score of the severity of the neurodegeneration.

The normative z-maps proposed here carry additional information with respect to the original images, because they directly encode the relationship between the single image and the average brain volumes in the reference population. This represents a great advantage in terms of the readability of the z-maps, where values closer to 0 indicate brain volumes closer to the expected pattern observed in the reference population. Indeed, observing for example a large TBM value does not indicate by itself whether the corresponding brain area has an ``outlying'' expansion that raises suspects of a disease, while through the z-maps this could be more easily assessed. 

The paper is structured as follows. We describe the features of the statistical model in \Cref{sec:methods}, with a focus on voxelwise distributions and prediction for new subjects. We also illustrate the computational aspects 
which avoid having to perform the estimation in all voxels while enforcing spatial smoothness. In \Cref{sec:appl}, we describe the TBM dataset used for the analysis and display the main features of interest (namely, a mean-variance relationship and skewness at the voxelwise level) which motivate our modelling choice. The results for the normative population are shown along with an example of z-map scoring by diagnosis group. Finally, we discuss potential further developments of the model in \Cref{sec:conclusions}.

\section{Methods}\label{sec:methods}
\subsection{Statistical framework}

Let $Y_i = \{Y_i(v), v \in \mathcal{V}\}$ be the brain image for the $i$-th subject ($i = 1,\dots, N$), whose domain is the closed cube $\mathcal{V} \in \mathbb{R}^3$. We assume that $Y_i$ is a square integrable random function on $\mathcal{V}$. In practice the domain $\mathcal{V}$ is discretised into $V$ voxels $v_1,\dots, v_V$, therefore in the next sections we will refer to the observed data for the $i$-th subject as ${Y_i(v_j),\;j=1,\dots,V}$.
%Consider a reference sample $Y_1, \dots, Y_N$. Let $Y_i$ be a realisation of a random function for the $i$-th subject ($i = 1,\dots,N$), with $Y_i = \{Y_i(v), v \in \mathcal{V}\}$.  We assume that $Y_i$ is a square integrable random function on the closed cube $\mathcal{V}$.
\par For $v \in \mathcal{V}$, define
\begin{equation}\label{eq:w_unif}
    U_i(v) = F_{\text{SN}}(Y_i(v); \mu(v), \sigma^{2}(v), \gamma(v)),
\end{equation}
where $F_{\text{SN}}$ denotes the skew-normal cumulative distribution function with mean function $\mu$, standard deviation function $\sigma$ and skewness function $\gamma)$ for the reference population \citep{azzalini2013skew, arellano2008centred}. The probability integral transform in \Cref{eq:w_unif} returns $U_i(v) \in (0,1), \; \forall v \in \mathcal{V}$: it is a latent uniform process based on the distribution $F_{\text{SN}}$ for the $i$-th subject. Covariates can be easily accommodated in this framework. For example, a linear effect of the covariates $X$ on the mean parameter can be included as $\mu(v) = \mathbf{X}\beta(v), \ \forall v \in \mathcal{V}$ (see \citealp{li2015csfm} for a more general formulation). In this work, we assume a linear model for the mean of the skew-normal distribution, with the design matrix $X$ is made of 4 columns: a vector of 1 for the intercept and the values of age, sex and their interaction.

The latent process $U_i$ has the role of incorporating the voxelwise distributional differences in a single object. At the voxelwise level, its interpretation as quantile of a distribution returns an immediate quantification of the distance from the mean of the population (identified as $U_i(v) = 0.5 \ \forall v \in \mathcal{V}$. The values of $U_i$ are therefore comparable between each other, in contrast to the original TBM values: indeed, while a TBM value of e.g. 1100 might be totally within a normal range in one brain location and extreme in another, a $U_i(v)$ value of 0.7 has the same interpretation across the whole brain.

Let $Y^*$ be a realisation of a random function for a new subject who could either belong or not to the reference population. We can compute the latent uniform process $U^*(v)$ using \Cref{eq:w_unif} with the estimated parameters obtained in the reference population, 
%the reference population to compute
%
%\begin{equation}\label{eq:latent_newobs_W}
%    U^*(v) = F_{\text{SN}}(Y^*(v); \mu(v), \sigma^{2}(v), \gamma(v)),
%\end{equation}
%
and set
\begin{equation}\label{eq:latent_newobs}
Z^*(v)=\Phi^{-1}(U^*(v)). 
\end{equation}
where $\Phi^{-1}$ is the inverse cumulative distribution function of a standard normal. This normative z-map gives information about how the image for the new subject compares to the reference population: values closer to zero indicate that the volume observed locally in the image for the subject is close to the mean value in the reference population, whereas more extreme values are potentially informative of non-healthy expansion/shrinkage of brain regions. The degree of information of the z-map is also more homogeneous across the brain than the original TBM values: while a value of 3 in the z-map indicates the same degree of extremity for all the voxels, a value of e.g.\ 1100 in a TBM image might be within a normative range in one brain area but not in another. The z-map for each subject could be therefore used to immediately locate those brain areas that depart from the mean of the normative population. 

Various scalar indices can be built to summarise the information carried by the z-maps into a single value. 
A simple approach is to plot first the z-values for each subject into a histogram and then take a summary statistics of the distributions or parts of it, like the tails. 
%Dealing with Gaussian voxelwise distributions, the easiest approach is to count the number (or proportion) of voxels for which the normative z-value is greater in absolute value than a certain threshold. For example, $n_3$ could be the number of voxels for which $\lvert Z(v) \rvert \geq 3$. A high value for this index suggests that in a relevant part of the brain the observed TBM values are very far from the mean in the reference population and therefore might show evidence of deviation from the healthy pattern.%This index could also be turned into a proportion by dividing for $V$, the total number of voxels.
%In this direction, we can consider an approach (described in \citealp{marquand2016understanding}) based on extreme value statistics. Each individual z-map is summarised by the (robust) mean of an extreme block (e.g.\ from the 99th percentile of the distribution of z-values for each subject). A parametric distribution from the generalised extreme value (GEV) family is then fitted on the normative sample using these averages, then for every subject the cumulative probability under the GEV distribution is used to quantitatively assess the extent of the deviation. 
For example, each individual z-map can be summarised by the (robust) mean of an extreme tail, e.g.\ from the 99th percentile of the distribution of z-values for each subject (other indices based on extreme value analysis are presented in \citealp{marquand2016understanding}). Under this approach, the specific application will drive the choice of the percentile defining the extreme tail (whether to deal with the highest, lowest, or both). In our setting, where the enlargement of the ventricles is balanced by the shrinkage of the cortex, it seems likely that extremes in both sides are carrying information, therefore we will define the summary $u_q^{abs},\ q \in (0,1)$ as the mean of the right tail above quantile $q$ of the distribution of absolute z-values.

%Further indication can be obtained from the histogram of the individual z-maps. By construction, the z-values are drawn from a standard normal distribution. %In the grid-based approach, this holds for the voxels of the grid and through smoothing for the rest of the brain, approximately. 
%A histogram for a subject belonging to the normative population would therefore be looking like a standard normal distribution too. Any departure from this distribution (e.g.\ in terms of variance or skewness) could be linked to a departure from the average in the normative population. In addition, evidence of multimodality or quantities like the test statistics for normality tests like Jarque--Bera, Anderson--Darling, Cramér--von Mises or Lilliefors could help in detecting non-Gaussian behaviours. 

\subsection{Model estimation}

In the literature about the skew-normal distribution \citep{azzalini2013skew, arellano2008centred} the set of parameters of $F_{\text{SN}}$ is called centred parameterisation (CP) to distinguish it from the original direct parameterisation (DP). The centred parameterisation is easier to interpret, as the parameters are functions of the first three moments in the population. CP is also the standard choice in estimation, because it removes the problem of singularity of the Fisher information matrix when the DP shape parameter (corresponding to the skewness parameter in CP) is equal to 0, which is linked to the violation of the asymptotic normality assumption of the maximum likelihood estimates \citep{azzalini2013skew}. The likelihood function for CP gets closer to a quadratic function and produces estimators which are less correlated than the DP estimators \citep{monti2003noteSN}. %An iterative procedure is needed to produce maximum likelihood estimates of the three parameters. %The sample moments can be used as starting points or the procedure. 

In \citet{azzalini2013skew} it is noted that the skewness parameter $\gamma$ is constrained within the set $(- c_\gamma, c_\gamma)$, with
\begin{equation}
    c_\gamma = \dfrac{\sqrt{2}(4 - \pi)}{(\pi - 2)^{3/2}} \approx 0.9953,
\end{equation}
while other distributions such as skew-$t$ might be more appropriate for larger observed sample skewness. The direction of the skewness is determined by the sign of $\gamma(v)$: if positive, the distribution is skewed to the right. For skewness equal to 0, SN reduces to a normal distribution with the same mean and variance. 

The maximisation of the likelihood of the skew-normal distribution is obtained via an iterative procedure. When covariates are available, a linear regression model for the location parameter is specified \citep{azzalini2013skew}. The data fitting could potentially be performed independently at each voxel in a parallel setting as the spatial dependence is captured at a later stage using a copula \citep{staicu2012skewedFDA}. Nevertheless, this approach is not ideal for two important reasons. The first reason is a computational one: although parallelisable, the number of voxels is still large enough to slow down the calculation of the parameters at the voxelwise level. The second reason is more conceptual: the full voxelwise approach will not ensure the smoothness of the functional parameter (which is a desirable outcome, because the noise induced by the discretisation is not of interest and because the storage in memory of the basis expansion is more efficient). 

\begin{figure}[hbtp]
    \centering
    \includegraphics[scale = 0.6]{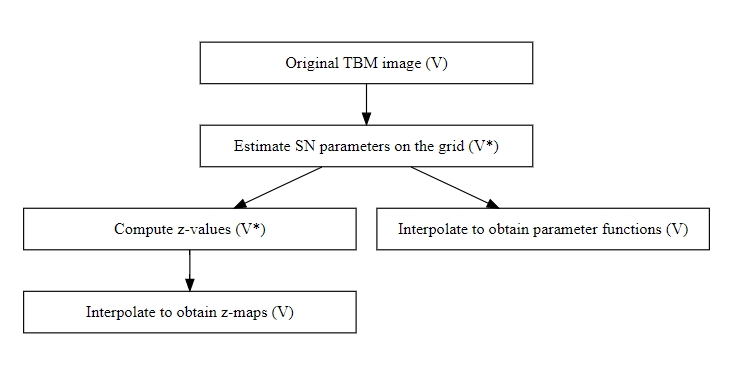}
    \caption{Workflow for the analysis on the grid (the number of voxels considered in the corresponding step is reported in brackets). Starting from the original TBM image, for each voxel in the grid the parameters of the skew normal distribution are estimated (and interpolated across the whole brain to obtain the parameter functions). Using those parameters, the z-values are computed on the grid. Radial basis functions (RBF) are then used to estimate the z-values in the rest of the brain image. Once the full z-maps are calculated for each subject, these can be used for further analysis (e.g.\ performing functional principal component analysis to identify the main modes of variation.}
    \label{fig:grid_computation_workflow}
\end{figure}

To bypass the need of running the computation for every voxel, we consider a subset of the voxels arranged in a regular grid $\mathcal{K}$ of size ($V^* \ll V$).
From now on, those selected voxels will be defined as $\{\kappa_1, \dots, \kappa_{V^*}\}$. For these voxels, the maximum likelihood estimates of the parameters are computed. For the $i$-th subject in the reference sample, the latent process $U_i$ and subsequently the standard Gaussian z-values can be computed as in \Cref{eq:w_unif} and \Cref{eq:latent_newobs}. The analysis on the grid (summarised in \Cref{fig:grid_computation_workflow}) is detailed below.

For the interpolation of the z-values for voxels outside the set $\mathcal{K}$, we use radial basis functions (RBF, \citealp{fasshauer2007choosing}). 
The number of radial basis functions is determined by the number of ``centres'', i.e.\ fixed points at which the radial function take their largest value. A basis function for each centre is computed, with non-zero weights in the voxels that are within a certain radius from the centre. The standard choice for centres is the same grid of preselected voxels $\{\kappa\}_{j=1}^{V^*}$ where we have carried the likelihood estimation out. %For the sake of simplicity, we recommend to use a regular grid, where the distance (in the 3 directions) is prespecified. This approach would guarantee that the distance between any voxels and the closest centre is within a certain range which depends on the grid spacing. Depending on the specific application, a grid with irregular spacing as in Chebyshev discretisation could also be chosen to capture finer changes in some areas. 
We recall that the observed value of the function at the $j$-th centre is $Z(\kappa_j)$.

Following the mathematical procedure illustrated in \citet{carr2001reconstruction}, we define the interpolant $s$ as a real-valued function with constraints $s^{*}(\kappa_j) = Z(\kappa_j)$ (that is the observed value of the function at the $j$-th centre). To define $s^{*}$, we build the matrix
\begin{equation}
    \bm{G} =
\left(
  \begin{array}{cc}
  \bm{H}^* & \bm{1} \\
  \bm{1}^T & 0
\end{array} \right)
\end{equation}
where the $V^* \times V^*$ symmetric matrix $\bm{H}^*$ contains the evaluation of the radial basis function $h$ for any distance $d$ between any pair of centres 
\begin{equation}
    H^*_{jl} = h(d(\kappa_j,\kappa_l)) \qquad j,l= 1, \dots,V^*. 
\end{equation}
and $\bm{1}$ is the $V^*$-dimensional vector whose elements are equal to 1.
The problem is now phrased in terms of a linear system: we are interested in finding the $V^*$-dimensional vector $\bm{b}$ and the scalar $b_0$ such that 
\begin{equation}
    Z(\kappa_j) = b_0 + \sum_{l=1}^{V^*} b_{l}H^*_{jl}.
\end{equation}
with a sum-to-one constraint on the vector $\bm{b}$.
\begin{comment}    
%
\begin{equation}
    \bm{G}
  \begin{bmatrix}
  \bm{b} \\
  b_0
\end{bmatrix} = 
\begin{bmatrix}
  Z(\bm{\kappa}) \\
  0
\end{bmatrix},
\end{equation} 
%
or analogously 
%
\begin{equation}
    Z(\kappa_j) = b_0 + \sum_{l=1}^{V^*} b_{l}H^*_{jl}.
\end{equation}
%
\end{comment}
The solution is now used to predict a value for a generic voxel: % $\{v_j\}_{j=1}^V$: 
\begin{equation}
\left(
  \begin{array}{cc}
  \bm{H} & \bm{1} \\
\end{array} \right) 
\begin{bmatrix}
  \bm{b} \\
  b_0
\end{bmatrix}
\end{equation}
where $H$ is the $V \times V$ matrix with the radial basis function evaluated at each voxel.

\begin{comment}
%
\begin{equation}
    H_{jk} = h(d(v_j - \kappa_k)) \qquad j= 1,\dots,V;\; k= 1,\dots,V^*. 
\end{equation}
%  
\end{comment}

The gain in computational efficiency that stems from applying basis functions on the grid instead of using all the voxels in the brain comes at a price. First, the performance of smoothing basis functions relies on some tuning parameters (such as the standard deviation for Gaussian RBF) for which it is not easy to determine optimality criteria. The best parameter is often chosen by trial-and-error or ad-hoc solutions (\citealp{fasshauer2007choosing}). Indeed, when the basis covers a larger area, the interpolation matrix becomes less sparse, while in the opposite case
%when the basis covers a larger area the interpolation matrix will be , while when it is In addition, there is a trade-off between increasing the accuracy (at the cost of getting a higher condition number) and  
%: choosing a small value for the tuning parameter increases accuracy but also the condition number of the interpolation matrix. Furthermore, for large values of the tuning parameter 
the so-called ``bed-of-nails'' interpolant is obtained: the function sharply peaks at the centres but decreases to 0 elsewhere. In the 3D grid case, we suggest to select a value for the tuning parameter that is smaller than the distance between consecutive centres in the grid. %To overcome the problem of the shape parameter, polyharmonic splines have been proposed (CITE). 

Another aspect of interpolation using radial basis function and polynomials is Runge's phenomenon, i.e.\ the approximation errors further from the centres are larger at the boundary of the domain \citep{fasshauer2007choosing, boyd2010six}. In the 3D brain imaging setting, although the brain mask has irregular boundaries in the three dimensions, this issue is not likely to be relevant, especially when the grid spacing (and consequently the maximum distance between a voxel and the closest centre) is moderate.

\section{Results}\label{sec:appl}

%\subsection{Mapping the TBM CN population}
%\subsection{Comparing individual images with the mean of the CN population}

\subsection{Data}

We build the normative model on a dataset from the Alzheimer's Disease Neuroimaging Initiative (ADNI), which consists of 817 adults (with age ranging between 54.4 and 90.9 years). A diagnosis is available for each of them: 229 subjects were considered as cognitively normal (CN), whereas 400 subjects were showing mild cognitive impairment (MCI) and 188 were diagnosed with Alzheimer's Disease.  %The sample used in \citet{palma2020quantifying} is a subset of the data analysed in this work.
\begin{table}[hbtp]
    \centering
    \begin{tabular}{c|cccc}
            \hline
        Diagnosis & Subjects & Proportion of females & Age mean & Age IQR\\
        \hline
         CN & 229 & 0.48 & 75.87 & 6.20 \\
         MCI &  400 & 0.36 & 74.74 & 10.23\\
         AD & 188 & 0.47 & 75.36 & 10.58\\
         \hline
    \end{tabular}
    \caption{Demographic characteristics of the subjects in the dataset. IQR: interquartile range.}
    \label{tab:table1}
\end{table}

The imaging data used in this work are tensor-based morphometry images (TBM). In a cross-sectional setting, each MRI scan is aligned to the \textit{minimal deformation template} (MDT) obtained by averaging several structural MRI scans \citep{hua2013unbiased}. The deformation induced by this alignment is mathematically described by a function that maps a 3-dimensional point in the template to the corresponding one in the individual image. The Jacobian matrix of the deformation incorporates the volume differences with respect to the MDT in terms of shearing, stretching and rotation. Its determinant evaluated at each voxel is a summary of local relative volumes compared to the MDT. Further details about TBM data are available in \citet{ashburner2004morphometry}.

A 3D preprocessed TBM image taken at the entrance of the ADNI study is available for each individual in the sample. The dimensions of the images are $220\times220\times220$, with voxel size equal to 1 mm$^3$. In the dataset, the threshold value is set at 1000: larger values indicate that expanded volume with respect to the MDT is observed in that specific voxel (lower values indicate shrinkage).

The mask used to subset only the part of the image that displays the brain is built with the same characteristics as described in \cite{palma2020quantifying}: we use a Gaussian kernel with standard deviation equal to 2 voxels (FWHM 4.7 mm) and threshold it at 0.5. Each masked image is made of approximately 2 million nonzero voxels. %(details on how the Gaussian kernel is used in imaging are reported in \citealp{poldracknichols2011handbook}). This reduces each image to a collection of more than 2 million voxels (approximately $20\%$ of the original dimensions).

\subsection{Exploratory analysis}

\begin{figure}[!p]
    \centering
    \includegraphics[width=0.455\linewidth]{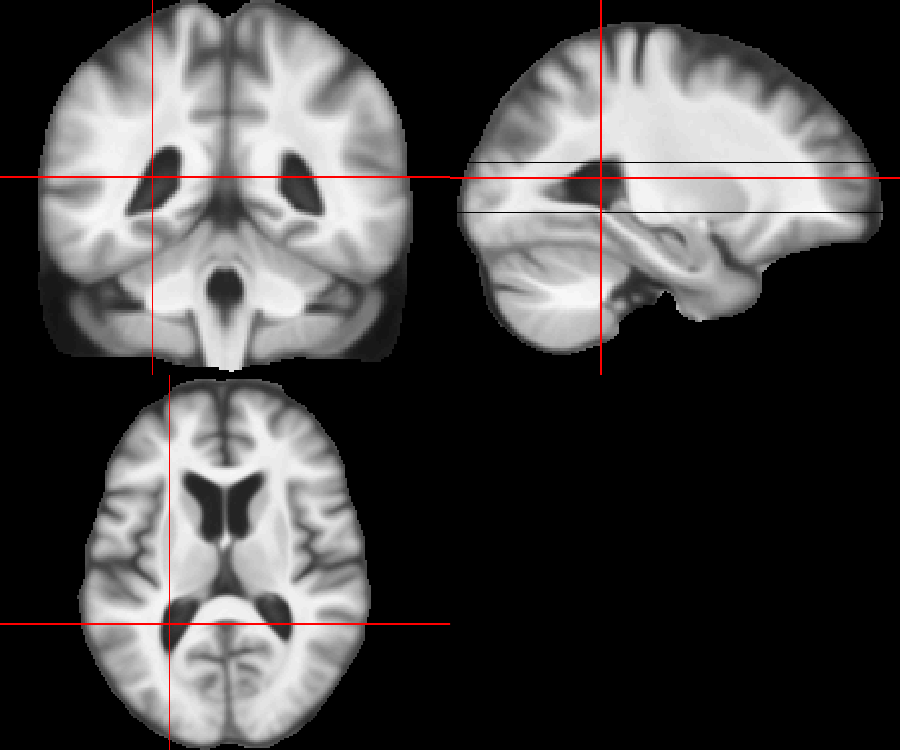}\hfill
    \includegraphics[width=0.53\linewidth]{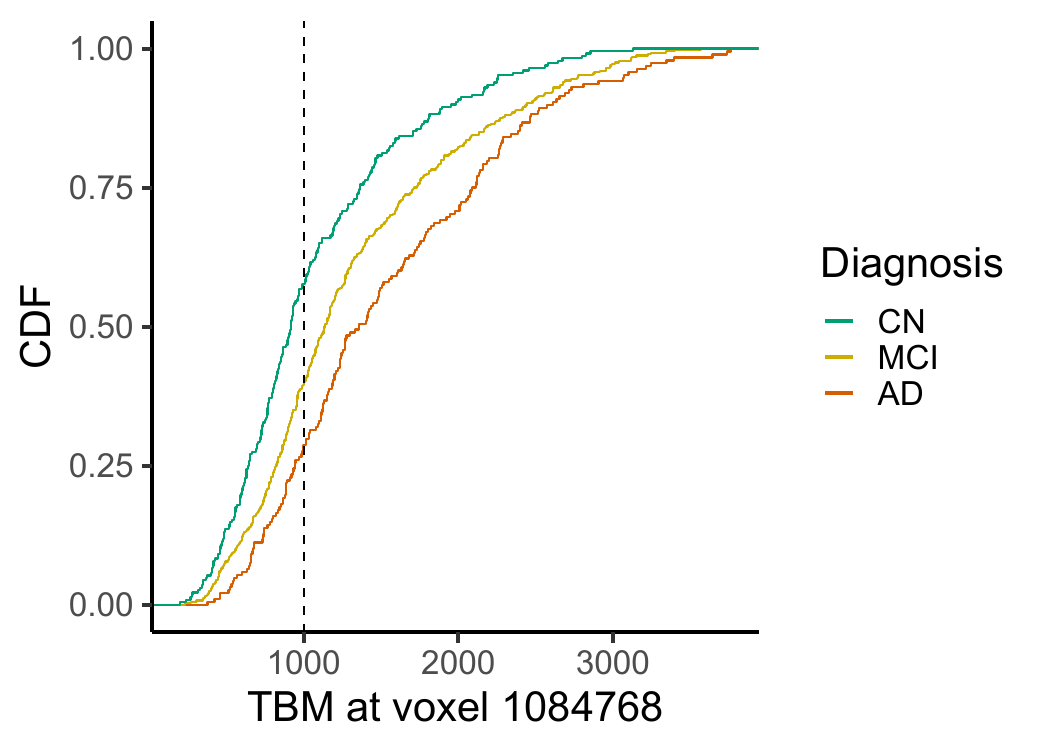}\\%motiv_A2long.pdf
    \includegraphics[width=0.455\linewidth]{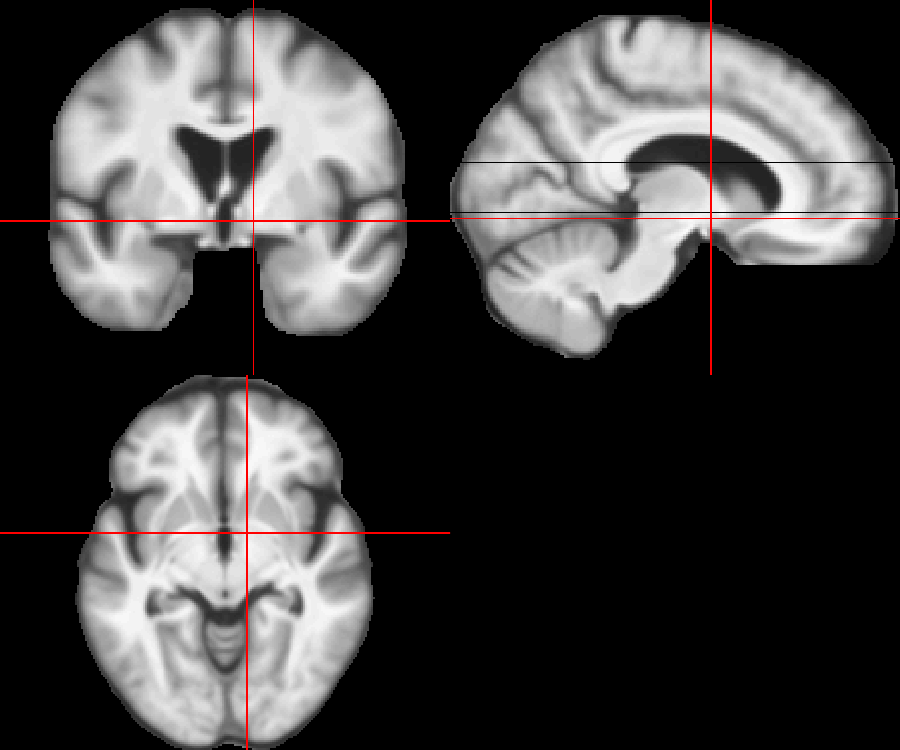}\hfill
    \includegraphics[width=0.53\linewidth]{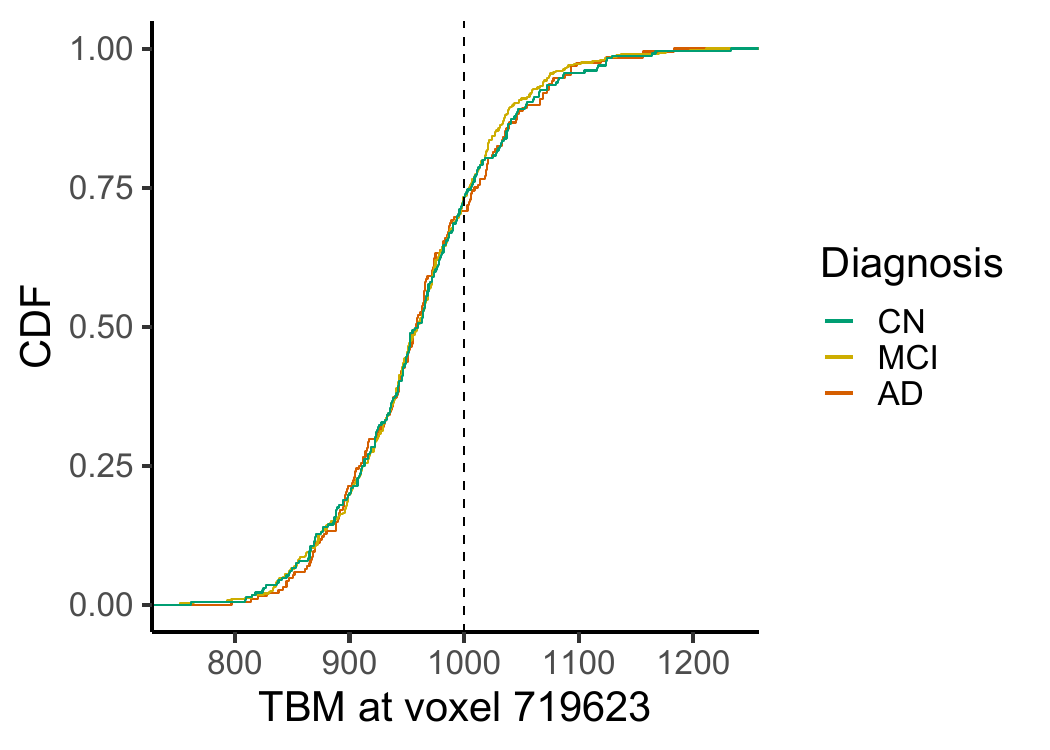}%motiv_B2long.pdf
    \caption{Top: empirical cumulative distribution functions of TBM Jacobian values by diagnosis group (right) for a voxel in the lateral ventricles. Bottom: empirical cumulative distribution functions of TBM Jacobian values by diagnosis group (right) for a voxel outside the lateral ventricles.}
    \label{fig:motivation_distrib}
\end{figure}

The exploratory analysis of the TBM images reveals spatial heterogeneity in the voxelwise distributions. \Cref{fig:motivation_distrib} shows the empirical cumulative distributions of TBM values by diagnosis groups for two example voxels in the brain, located inside and outside the lateral ventricles, respectively. The patterns observed are very different: for the voxel in the ventricles, the probability of observing more extreme values is larger for AD compared to MCI and CN, while the voxel outside the ventricles shows no clear differences between the cumulative distributions across the groups. %Furthermore, in the voxel within the ventricles, the distributions for all the groups are not symmetric.

\begin{sidewaysfigure*}
    \centering
    \includegraphics[width=0.99\textwidth]{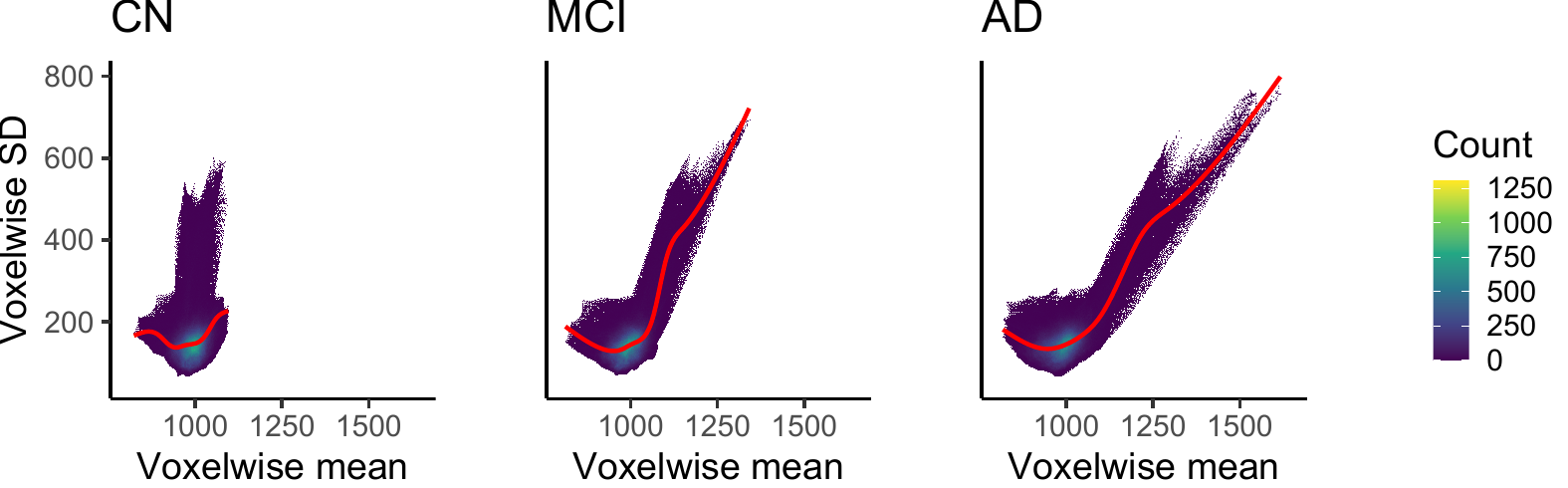}
    \caption{2D histograms of voxelwise means and standard deviations (SD) by diagnosis group. The number of bins is fixed to 600. A smooth regression line (obtained via penalised cubic splines) is added in red.}
    \label{fig:meanSDplots}
\end{sidewaysfigure*}

\begin{figure}
    \centering
    \includegraphics[scale = 0.75]{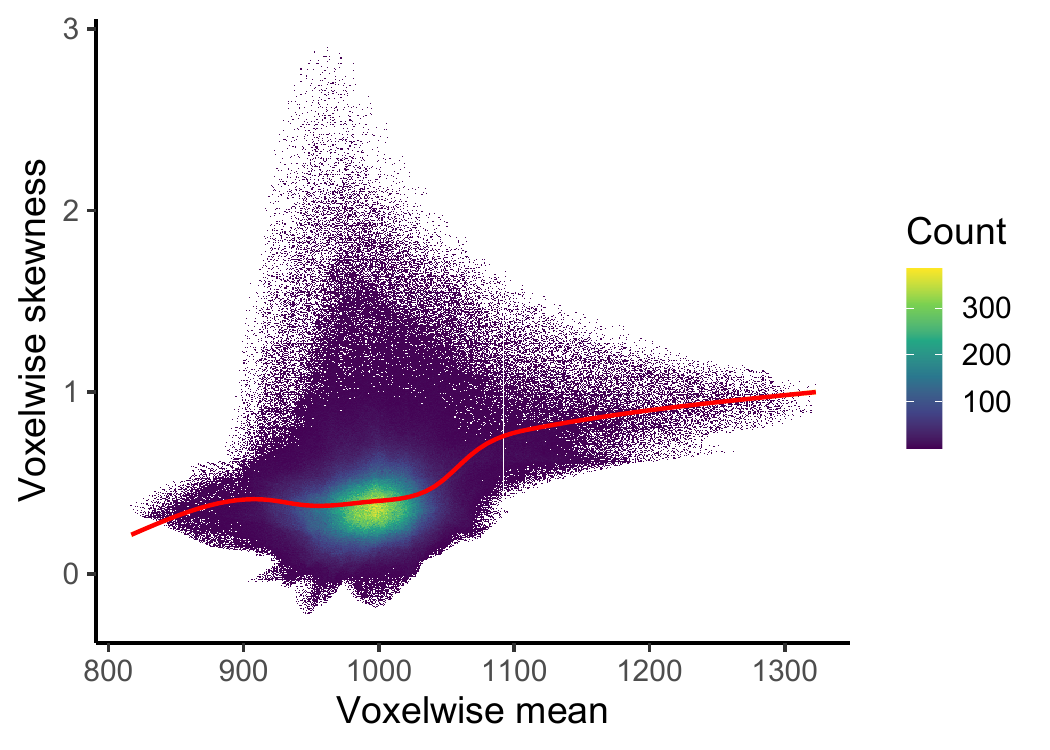}
    \caption{2D histogram of the voxelwise mean and skewness across all subjects. The number of bins is fixed to 600. A smooth regression line (obtained via penalised cubic splines) is added in red.}
    \label{fig:summaryplots_all}
\end{figure}

When looking at the summary statistics for all the voxels across all subjects, the patterns between diagnosis groups are even more evident. \Cref{fig:meanSDplots} shows the relationship between voxelwise means and standard deviations for each group. While most voxels show a mean around 1000, larger mean and variances are observed for some voxels, especially for the groups with diseases. But even for the cognitively normal subjects, the variability of the standard deviations changes as the means increase. When we compute the coefficient of skewness on the whole dataset (\Cref{fig:summaryplots_all}), brain areas with larger mean (e.g.\ the lateral ventricles) tend to exhibit more skewed distributions. 

%These considerations\todoDONE{Remove plot with skewness - change caption} show that the characteristics of the voxelwise distributions are highly heterogeneous across the brain and could play a role in the statistical models for brain images. The statistical properties of the Jacobian values in TBM have been studied in the literature. 

\subsection{Mapping the cognitively normal population}

We consider first the normative sample (training set) to compute the skew-normal parameter estimates: it is made of 183 CN subjects (approximately 80$\%$ of the CN group), selected after stratification by age group and sex (for a total of 95 males and 88 females). We carry out the skew-normal fitting procedure on a regular grid with 8mm spacing in the three dimensions. This returns 3949 voxels, approximately equal to $0.2\%$ of all the voxels within the mask. For these voxels, the skew-normal likelihood optimisation (with centred parameterisation) is carried out using the \textsf{R} package \textsf{sn} \citep{snpackage}. %The procedure takes approximately 30 seconds in a standard laptop in a serial version.

Radial basis functions (RBF) with Gaussian kernel and standard deviation $\varepsilon = 5.33$ mm ($66.67\%$ of the grid spacing) are used to interpolate the SN parameter functions across the rest of the brains. %For other values below 8mm tried on the same dataset, we do not observe the bed-of-nails problem, while for higher grid spacing the interpolation quality is poor. 
This value represented a compromise to achieve good interpolation quality while avoiding the bed-of-nails issue. A tensor product with univariate B-splines (as the one in \citealp{palma2020quantifying}) spaced every 8mm has been also used on the same training set but some analyses (not shown here) indicate that it is less convenient in terms of memory storage %that the number of B-splines functions is higher than for RBF (10920 against 3949 basis functions) and seem to suggest also that they 
and does not improve the quality of the interpolation, especially at the boundaries of the mask where approximation errors are larger. In this procedure, fitting the skew normal parameters on the grid takes approximately 3 minutes on a standard laptop. Turning the original brain scans into z-maps based on the parameter function takes approximately 1 minute per image: this step can be run in parallel.

\begin{sidewaysfigure*}
    \centering
    \includegraphics[width=0.33\textwidth]{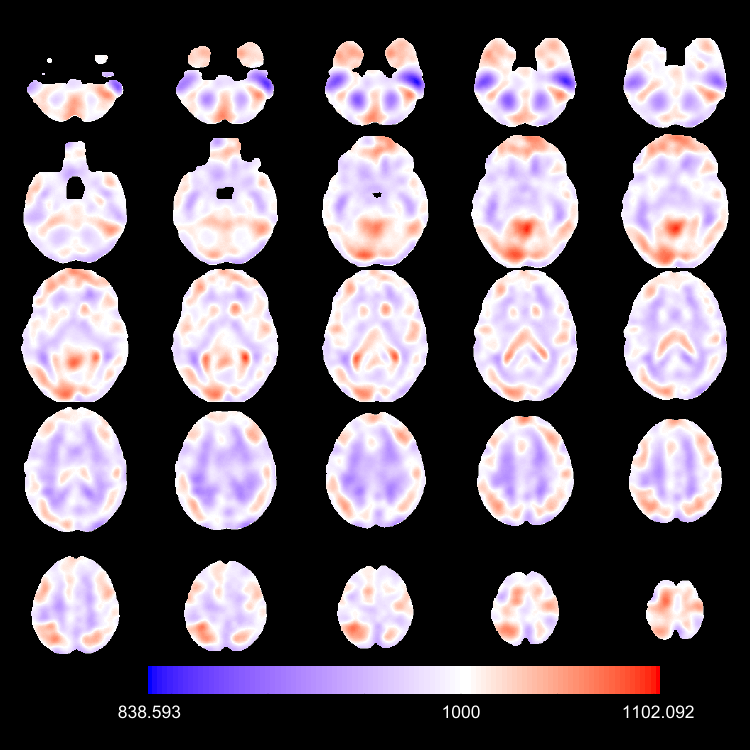}\hfill
    \includegraphics[width=0.33\textwidth]{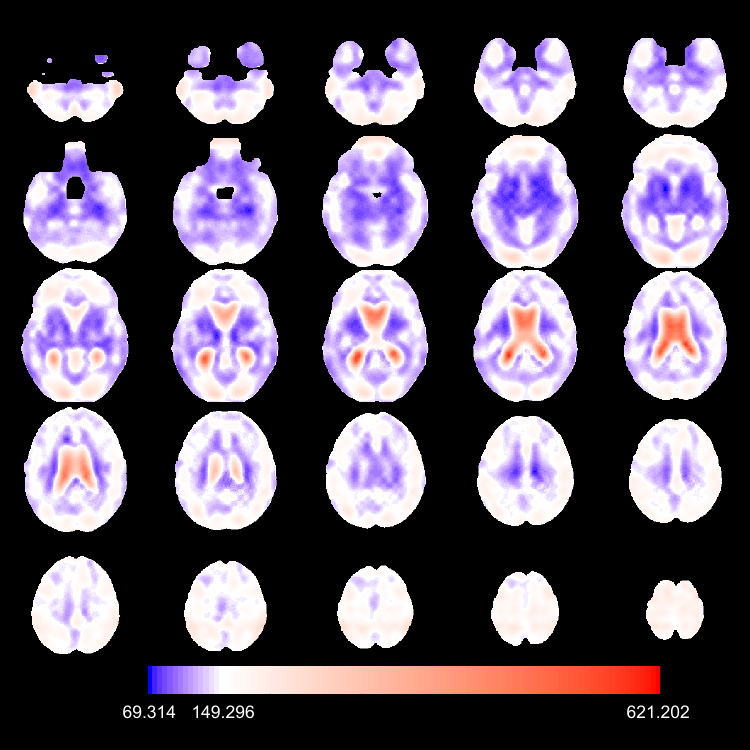}\hfill
    \includegraphics[width=0.33\textwidth]{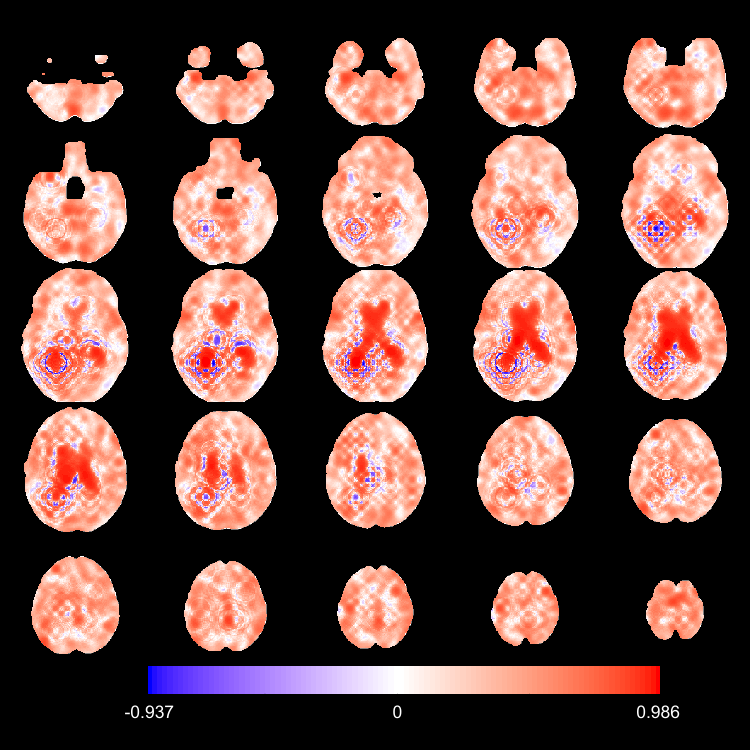}
    \caption{Axial slices of the mean (left), standard deviation (centre) and skewness (right) parameter functions from skew-normal fitting in the normative population. Slices are ordered from bottom to top. For the standard deviation, the colour white corresponds to the average standard deviation in the normative population.}
    \label{fig:SNparamfunctions}
\end{sidewaysfigure*}

The parameter functions are plotted in \Cref{fig:SNparamfunctions}. The mean and the standard deviation are larger in the lateral ventricles than the rest of the brain. The estimated skewness is greater than 0 across almost the whole mask. 

\begin{sidewaysfigure*}
    \centering
    \includegraphics[width=0.33\textwidth]{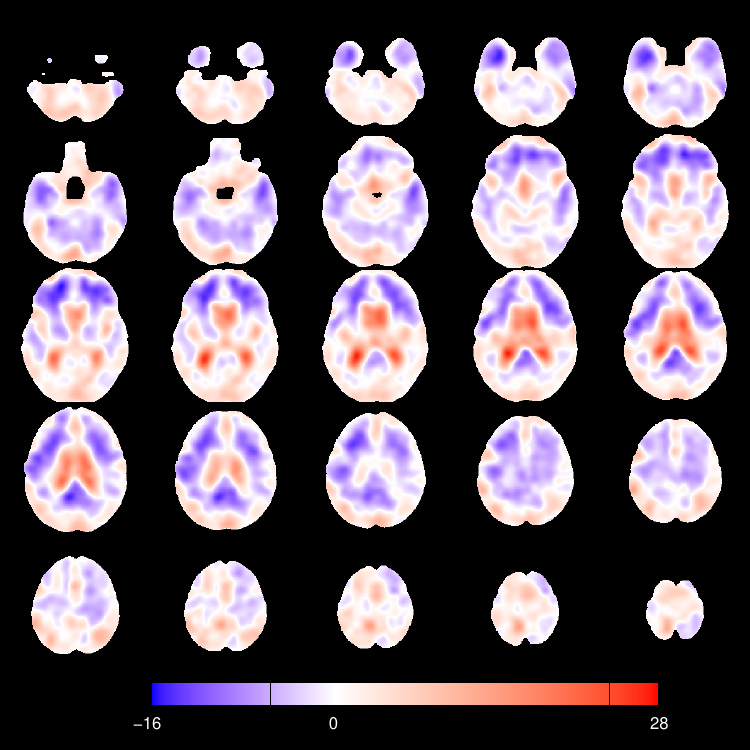}%\hfill
    \includegraphics[width=0.33\textwidth]{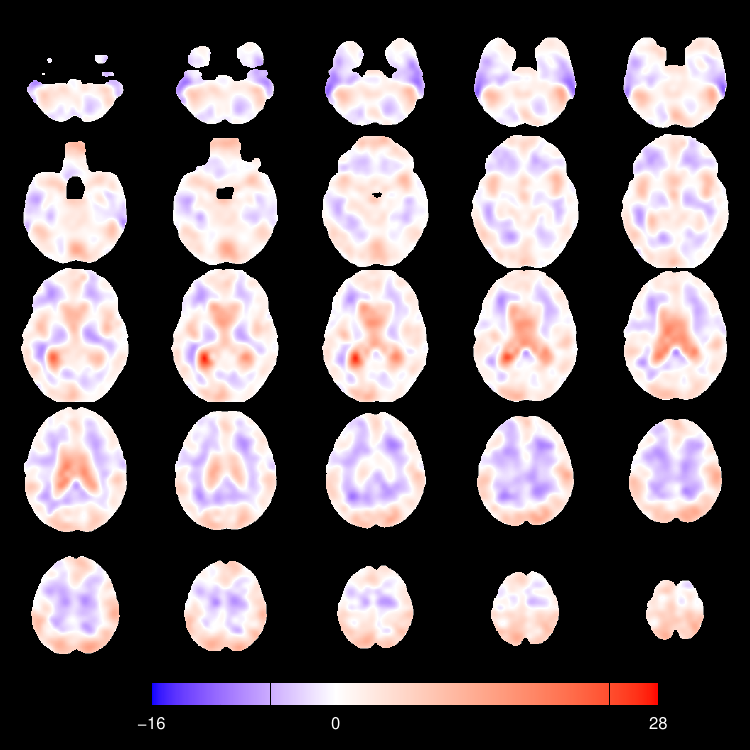}
    \caption{Axial slices of the age association with the mean function for females (left) and males (right) in the normative population. Slices are ordered from bottom to top.}
    \label{fig:SNmean_covariates}
\end{sidewaysfigure*}

For the mean of the normative population, we also display the linear effects of age and sex (and their interaction) on the TBM maps (\Cref{fig:SNmean_covariates}). The mean TBM values (not shown here) for males are larger in the lateral ventricles and in the top part of the brain. A one-year change in age for females is also linked to an expansion in the lateral ventricles, balanced by the shrinkage in other parts of the brain, and in particular in the frontal lobe. The same one-year change in age for males mostly affects the same brain regions, although the volume changes are more modest\todo{sex effect has bigger legend}. 

\begin{sidewaysfigure*}
    \centering
    \includegraphics[width=0.33\textwidth]{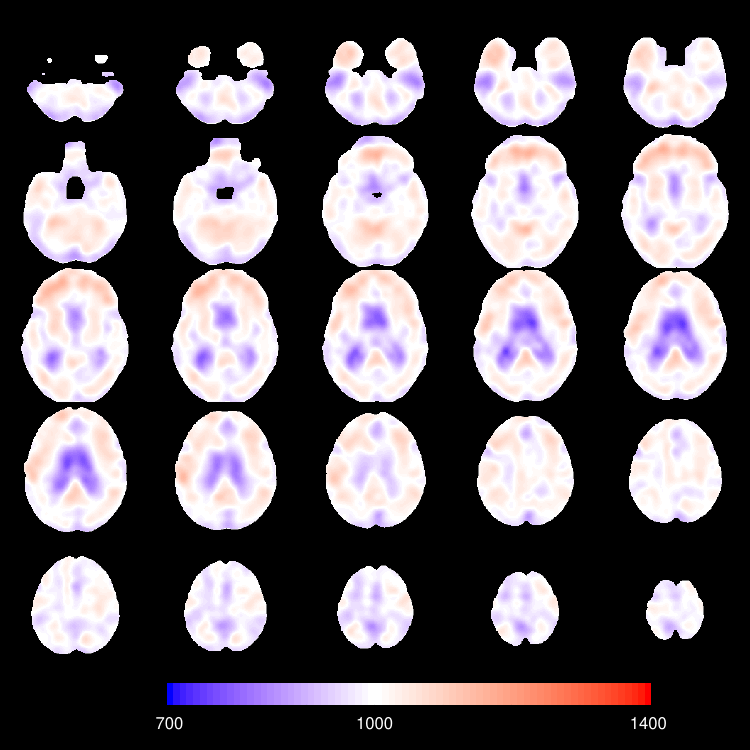}\hfill
    \includegraphics[width=0.33\textwidth]{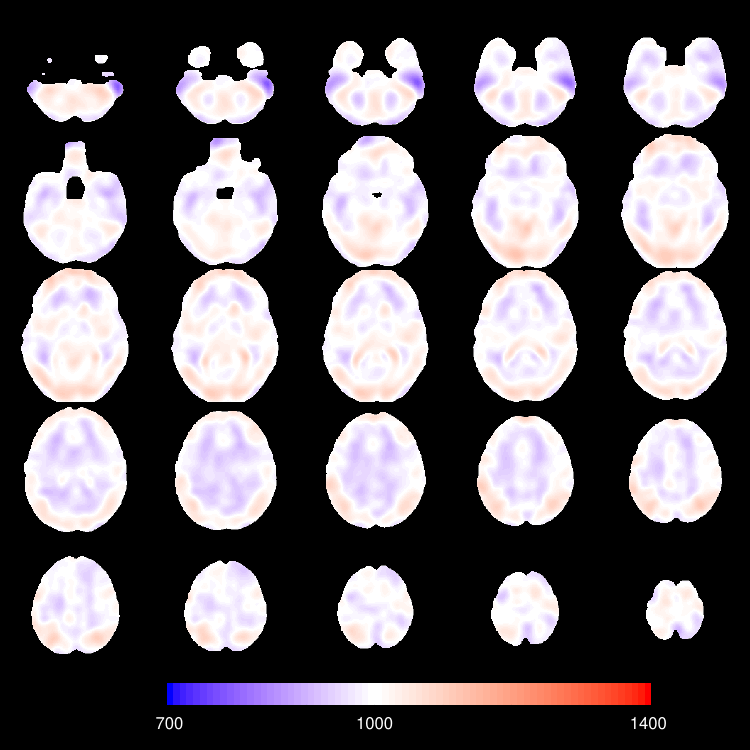}\hfill
    \includegraphics[width=0.33\textwidth]{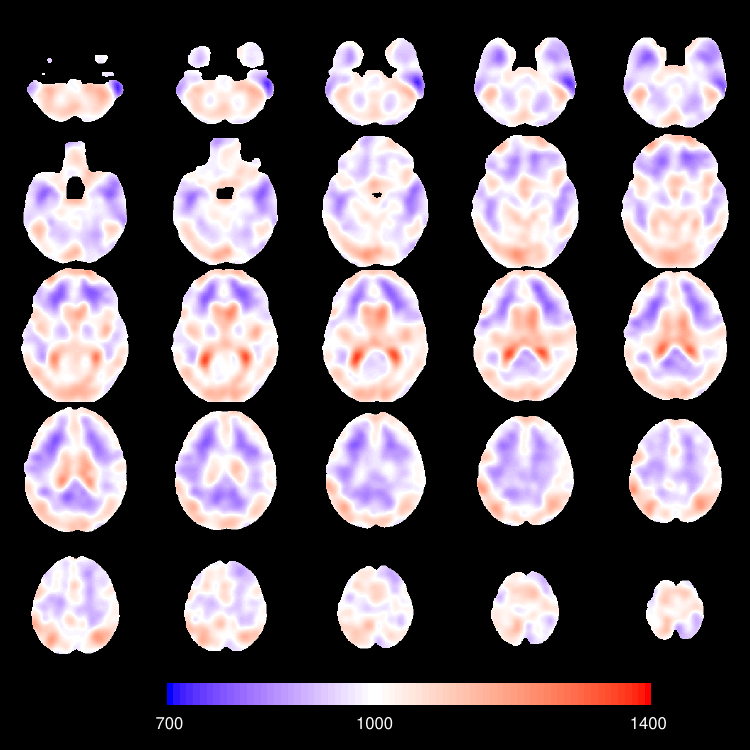}
    \caption{Axial slices of the mean for females at 70 years (left), 80 years (centre) and 90 years (right) in the normative population. Slices are ordered from bottom to top.}
    \label{fig:SNmean_femalepred}
\end{sidewaysfigure*}

These functional coefficients are used to obtain mean predictions for specific covariate values in the normative population. As an example, the mean for females at 70, 80 and 90 years are displayed in \Cref{fig:SNmean_femalepred}. While at 70 years old the lateral ventricles are smaller than the TBM template, their volume becomes bigger than the template at 80 and increases at 90 years old. An opposite trend is observed for the volume in the frontal regions. In terms of intensity of the TBM values, more extreme values are observed at the highest age level.

\begin{figure}[!p]
    \centering
    \includegraphics[scale = 0.8]{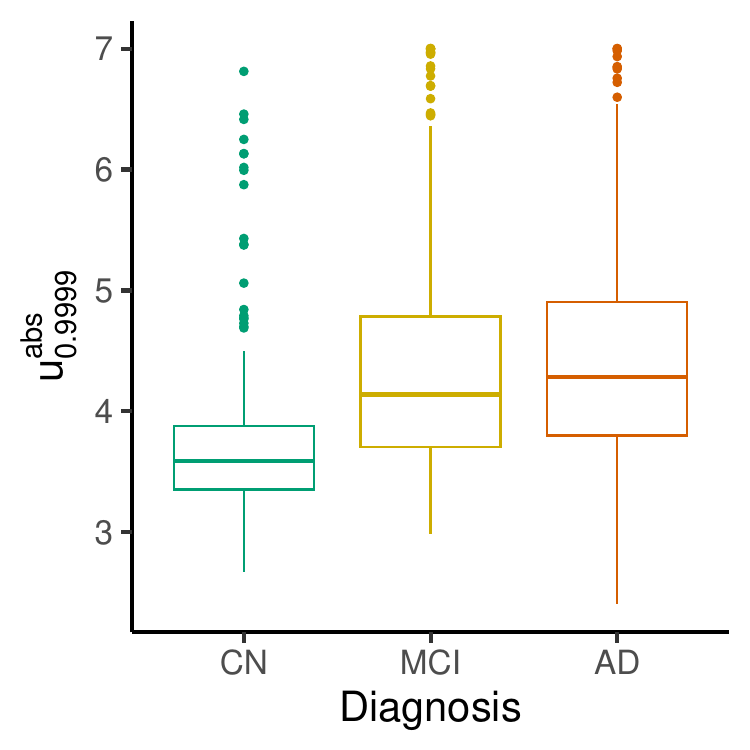}\hfill
    \includegraphics[scale = 0.8]{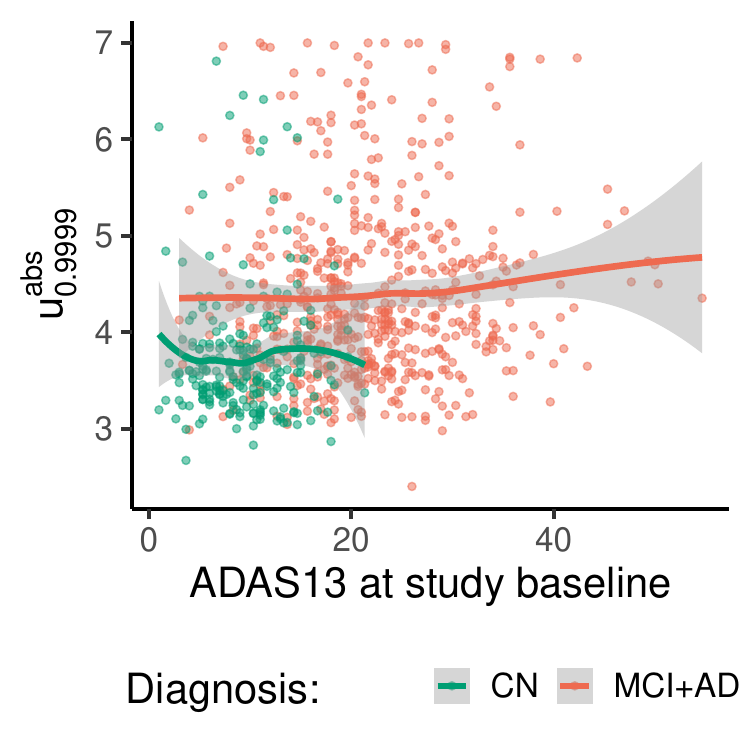}
    \caption{Top: Boxplots of $u_{0.9999}^{abs}$ by diagnosis group. Bottom: Plot of $u_{0.9999}^{abs}$ by ADAS13 and diagnosis group.}
    \label{fig:uabs}
\end{figure}

%\subsection{Comparing individual images with the mean of the CN population}
The z-maps computed using the skew-normal parameter values at the grid and then smoothed across the rest of the brain are obtained for both the subjects in the normative population and the test sets. \Cref{fig:uabs} (top) shows the boxplots of $u_{0.9999}^{abs}$, the index of deviation obtained by averaging the top 0.01$\%$ z-values in absolute values for each subject. Some evidence of a trend between this index and the severity of disease status is observed, although no relevant discrimination across groups is observed. In addition, when the index of deviation is plotted against ADAS13 \citep{kueper2018adas}, a neuropsychological test often used in AD studies to assess cognitive dysfunctions (the larger the score, the higher is the disease severity),
the regression lines for the diseased groups remain above the one for cognitively normal subjects (\Cref{fig:uabs}, bottom). Similar results (not reported here) were obtained for different thresholds of the right tail of the distribution.

\newpage
\section{Conclusions and further developments}\label{sec:conclusions}

The analysis of brain morphometry images is of large interest due to its ability to show and quantify signs of atrophy within different brain regions. Using a dataset from ADNI, we have shown that the mean, standard deviation and skewness of the voxelwise distributions of TBM values exhibit interesting patterns. In this work we have proposed a methodology to take into account these characteristics by using a skew-normal distribution at the voxel level, proposing a normative model to study brain volumes in absence of neurodegeneration. The normative approach provides then a set of reference parameters on which to build individual brain maps, which can then be summarised into single indices. By using this approach, we aim at identifying subjects with cognitive impairment as ``extremes'' with respect to the reference population, by computing individual risk scores. 

A novel outcome of the normative model is the quantification and visualisation of heterogeneity across brain regions within the healthy population, which is not usually obtained in a case-control setting. In particular, the variance parameter function shows that even in the cognitively normal group the voxel with the highest variability are located within the lateral ventricles. This atlas of TBM variability could be useful to better understand age-related effects in regional volume changes and help disentangling pathological degeneration from the normal age effect.

Another novel outcome is the easy-to-implement procedure to reference a test image to the normative population. The calculation of the z-maps requires only the parameter functions. This evaluation is therefore more appropriate when the identification of abnormal volumes is of interest. In contrast with other normative models, with our approach we openly enforce smoothness of the z-maps.

A strength of this approach is the large flexibility given by the modular structure of the workflow. In this analysis, a single distribution family (the skew-normal) was chosen as it provided enough flexibility to deal with the spatial heterogeneity between voxels within and outside the lateral ventricles. Nevertheless, other continuous distributions could be explored; information criteria such as AIC and BIC can inform the choice between multiple candidate distributions \citep{bethlehem2022brain}. The indices of deviations can be also tailored for specific interests (e.g.\ considering only positive or negative extremes). In addition, other quantities like functional data depth \citep{lopez2009depth,mosler2012depth,Gijbels2017depth} could be extended to the case of 3D imaging data and applied in this context.

In addition, the entire workflow might be used also used as a preliminary step for further analysis based on a normative population. Normative z-maps could be used for example within a regression framework to predict the disease status or transition from cognitively normal to AD. For these tasks, scalar-on-function regression \citep{morris2015functional} employs the whole function as the independent variable, for example through its basis representation or the scores obtained from a functional principal component analysis \citep{RamSil2005}.

A limitation of the model is the quantification of the spatial dependence between voxels. By smoothing the z-maps and the parameter functions we effectively introduce some dependence between the voxelwise distributions but a full account of the spatial information would require to use a copula as in \citet{staicu2012skewedFDA}. In the 3D functional data setting this object is difficult to visualise, but functional principal component analysis (see \citealp{palma2020quantifying} for the 3D implementation in brain imaging) provides a decomposition which can identify the brain areas where the variability between individuals is higher. A preliminary analysis (not reported here) confirm that there is large variability in the lateral ventricles even after the voxelwise heterogeneity is taken into account via the z-maps. In the ADNI application, the study of the copula might help to uncover spatial correlation patterns typical of healthy individuals and differentiate individuals from AD from the normative population. 

Finally, this approach could be extended beyond the application described in this work. For example, in a longitudinal framework the normative approach could be used to track the evolution of both the normative population and the subject-specific map. Furthermore, the procedure can be generalised to deal with other types of images or other neuroimaging-related curves, as well as to provide a potential framework for 3D data simulation in the reference population.

\newpage

\noindent \textbf{Declarations of interest}\\
All authors declare no conflict of interests.

\medskip

\noindent \textbf{Acknowledgements and funding sources}\\
MP is currently funded by the MRC grant MR/V020595/1. A substantial part of the work has been also funded by the EPSRC and MRC Centre for Doctoral Training in Next Generation Statistical Science: The Oxford-Warwick Statistics Programme (EP/L016710/1). TEN is supported by the Wellcome Trust, 100309/Z/12/Z.\\ We would like to thank David Firth, Xavier Didelot, Jeff Goldsmith %and the anonymous reviewers
for their insightful comments about the work. We also thank Paul Thompson and Xue Hua for the TBM data. 

\medskip

\noindent \textbf{Data availability statement}\\
Data are available under acceptance of the Data Use Agreement through the LONI Image and Data Archive (IDA). Data collection and sharing for this project was funded by the Alzheimer's Disease Neuroimaging Initiative (ADNI) (National Institutes of Health Grant U01 AG024904) and DOD ADNI (Department of Defense award number W81XWH-12-2-0012). ADNI is funded by the National Institute on Aging, the National Institute of Biomedical Imaging and Bioengineering, and through generous contributions from the following: AbbVie, Alzheimer’s Association; Alzheimer’s Drug Discovery Foundation; Araclon Biotech; BioClinica, Inc.; Biogen; Bristol-Myers Squibb Company; CereSpir, Inc.; Cogstate;Eisai Inc.; Elan Pharmaceuticals, Inc.; Eli Lilly and Company; EuroImmun; F. Hoffmann-La Roche Ltd and its affiliated company Genentech, Inc.; Fujirebio; GE Healthcare; IXICO Ltd.; Janssen Alzheimer Immunotherapy Research \& Development, LLC.; Johnson \& Johnson Pharmaceutical Research \& Development LLC.; Lumosity; Lundbeck; Merck \& Co., Inc.; Meso Scale Diagnostics, LLC.; NeuroRx Research; Neurotrack Technologies;Novartis Pharmaceuticals Corporation; Pfizer Inc.; Piramal Imaging; Servier; Takeda Pharmaceutical Company; and Transition Therapeutics. The Canadian Institutes of Health Research is providing funds to support ADNI clinical sites in Canada. Private sector contributions are facilitated by the Foundation for the National Institutes of Health (\url{www.fnih.org}). The grantee organization is the Northern California Institute for Research and Education, and the study is coordinated by the Alzheimer’s Therapeutic Research Institute at the University of Southern California. ADNI data are disseminated by the Laboratory for Neuro Imaging at the University of Southern California.

\medskip

\noindent \textbf{Author contributions (CRediT)}\\
Conceptualisation: all authors. Data acquisition: MP, TEN. Formal analysis and visualisation: MP. Writing original draft: MP. Writing – review \& editing: all authors.

\bigskip
\newpage
\bibliographystyle{apalike}
\bibliography{skewedFDA}

\begin{thebibliography}{}

\bibitem[Arellano-Valle and Azzalini, 2008]{arellano2008centred}
Arellano-Valle, R.~B. and Azzalini, A. (2008).
\newblock The centred parametrization for the multivariate skew-normal
  distribution.
\newblock {\em Journal of multivariate analysis}, 99(7):1362--1382.

\bibitem[Ashburner and Friston, 2004]{ashburner2004morphometry}
Ashburner, J. and Friston, K. (2004).
\newblock Morphometry.
\newblock In Frackowiak, R. S.~J., Friston, K.~J., Frith, C.~D., Dolan, R.~J.,
  Price, C.~J., Zeki, S., Ashburner, J.~T., and Penny, W.~D., editors, {\em
  Human Brain Function}, chapter~36, pages 707--722. Elsevier, 2 edition.

\bibitem[Azzalini, 2013]{azzalini2013skew}
Azzalini, A. (2013).
\newblock {\em The skew-normal and related families}, volume~3.
\newblock Cambridge University Press.

\bibitem[Azzalini, 2020]{snpackage}
Azzalini, A. (2020).
\newblock {\em The {R} package \texttt{sn}: The Skew-Normal and Related
  Distributions such as the Skew-$t$ (version 1.5-5).}
\newblock Universit\`a di Padova, Italia.

\bibitem[Bethlehem et~al., 2022]{bethlehem2022brain}
Bethlehem, R.~A., Seidlitz, J., White, S.~R., Vogel, J.~W., Anderson, K.~M.,
  Adamson, C., Adler, S., Alexopoulos, G.~S., Anagnostou, E., Areces-Gonzalez,
  A., et~al. (2022).
\newblock Brain charts for the human lifespan.
\newblock {\em Nature}, pages 1--11.

\bibitem[Boyd, 2010]{boyd2010six}
Boyd, J.~P. (2010).
\newblock Six strategies for defeating the {R}unge phenomenon in {G}aussian
  radial basis functions on a finite interval.
\newblock {\em Computers \& Mathematics with Applications}, 60(12):3108--3122.

\bibitem[Carr et~al., 2001]{carr2001reconstruction}
Carr, J.~C., Beatson, R.~K., Cherrie, J.~B., Mitchell, T.~J., Fright, W.~R.,
  McCallum, B.~C., and Evans, T.~R. (2001).
\newblock Reconstruction and representation of 3{D} objects with radial basis
  functions.
\newblock In {\em Proceedings of the 28th annual conference on Computer
  graphics and interactive techniques}, pages 67--76.

\bibitem[Chen et~al., 2015]{chen2015quantile}
Chen, H., Kelly, C., Castellanos, F.~X., He, Y., Zuo, X.-N., and Reiss, P.~T.
  (2015).
\newblock Quantile rank maps: a new tool for understanding individual brain
  development.
\newblock {\em Neuroimage}, 111:454--463.

\bibitem[Chung, 2013]{chung2013computational}
Chung, M.~K. (2013).
\newblock {\em Computational neuroanatomy: The methods}.
\newblock World Scientific.

\bibitem[Chung et~al., 2003]{chung2003deformation}
Chung, M.~K., Worsley, K.~J., Robbins, S., Paus, T., Taylor, J., Giedd, J.~N.,
  Rapoport, J.~L., and Evans, A.~C. (2003).
\newblock Deformation-based surface morphometry applied to gray matter
  deformation.
\newblock {\em NeuroImage}, 18(2):198--213.

\bibitem[Fasshauer and Zhang, 2007]{fasshauer2007choosing}
Fasshauer, G.~E. and Zhang, J.~G. (2007).
\newblock On choosing “optimal” shape parameters for {RBF} approximation.
\newblock {\em Numerical Algorithms}, 45(1-4):345--368.

\bibitem[Gijbels and Nagy, 2017]{Gijbels2017depth}
Gijbels, I. and Nagy, S. (2017).
\newblock On a general definition of depth for functional data.
\newblock {\em Statistical Science}, 32(4).

\bibitem[Hua et~al., 2013]{hua2013unbiased}
Hua, X., Hibar, D.~P., Ching, C.~R., Boyle, C.~P., Rajagopalan, P., Gutman,
  B.~A., Leow, A.~D., Toga, A.~W., Jack~Jr, C.~R., Harvey, D., Weiner, M.~W.,
  Thompson, P.~M., and the Alzheimer’s Disease Neuroimaging~Initiative
  (2013).
\newblock Unbiased tensor-based morphometry: improved robustness and sample
  size estimates for {A}lzheimer's disease clinical trials.
\newblock {\em Neuroimage}, 66:648--661.

\bibitem[Kueper et~al., 2018]{kueper2018adas}
Kueper, J.~K., Speechley, M., and Montero-Odasso, M. (2018).
\newblock The {A}lzheimer’s disease assessment scale--cognitive subscale
  ({ADAS-C}og): modifications and responsiveness in pre-dementia populations. a
  narrative review.
\newblock {\em Journal of Alzheimer's Disease}, 63(2):423--444.

\bibitem[Leow et~al., 2007]{leow2007statistical}
Leow, A.~D., Yanovsky, I., Chiang, M.-C., Lee, A.~D., Klunder, A.~D., Lu, A.,
  Becker, J.~T., Davis, S.~W., Toga, A.~W., and Thompson, P.~M. (2007).
\newblock Statistical properties of {J}acobian maps and the realization of
  unbiased large-deformation nonlinear image registration.
\newblock {\em IEEE transactions on medical imaging}, 26(6):822--832.

\bibitem[Li et~al., 2015]{li2015csfm}
Li, M., Staicu, A.-M., and Bondell, H.~D. (2015).
\newblock Incorporating covariates in skewed functional data models.
\newblock {\em Biostatistics}, 16(3):413--426.

\bibitem[L{\'o}pez-Pintado and Romo, 2009]{lopez2009depth}
L{\'o}pez-Pintado, S. and Romo, J. (2009).
\newblock On the concept of depth for functional data.
\newblock {\em Journal of the American Statistical Association},
  104(486):718--734.

\bibitem[Marquand et~al., 2019]{marquand2019conceptualizing}
Marquand, A.~F., Kia, S.~M., Zabihi, M., Wolfers, T., Buitelaar, J.~K., and
  Beckmann, C.~F. (2019).
\newblock Conceptualizing mental disorders as deviations from normative
  functioning.
\newblock {\em Molecular psychiatry}, 24(10):1415--1424.

\bibitem[Marquand et~al., 2016a]{marquand2016understanding}
Marquand, A.~F., Rezek, I., Buitelaar, J., and Beckmann, C.~F. (2016a).
\newblock Understanding heterogeneity in clinical cohorts using normative
  models: beyond case-control studies.
\newblock {\em Biological psychiatry}, 80(7):552--561.

\bibitem[Marquand et~al., 2016b]{marquand2016beyond}
Marquand, A.~F., Wolfers, T., Mennes, M., Buitelaar, J., and Beckmann, C.~F.
  (2016b).
\newblock Beyond lumping and splitting: a review of computational approaches
  for stratifying psychiatric disorders.
\newblock {\em Biological psychiatry: cognitive neuroscience and neuroimaging},
  1(5):433--447.

\bibitem[Monti et~al., 2003]{monti2003noteSN}
Monti, A.~C. et~al. (2003).
\newblock A note on the estimation of the skew normal and the skew exponential
  power distributions.
\newblock {\em Metron}, 61(2):205--219.

\bibitem[Morris, 2015]{morris2015functional}
Morris, J.~S. (2015).
\newblock Functional regression.
\newblock {\em Annual Review of Statistics and Its Application}, 2:321--359.

\bibitem[Mosler and Polyakova, 2012]{mosler2012depth}
Mosler, K. and Polyakova, Y. (2012).
\newblock General notions of depth for functional data.
\newblock {\em arXiv preprint arXiv:1208.1981}.

\bibitem[Palma et~al., 2020]{palma2020quantifying}
Palma, M., Tavakoli, S., Brettschneider, J., Nichols, T.~E., and ADNI (2020).
\newblock Quantifying uncertainty in brain-predicted age using scalar-on-image
  quantile regression.
\newblock {\em NeuroImage}.

\bibitem[Ramsay and Silverman, 2005]{RamSil2005}
Ramsay, J.~O. and Silverman, B.~W. (2005).
\newblock {\em Functional Data Analysis}.
\newblock Springer Series in Statistics. Springer.

\bibitem[Rutherford et~al., 2022]{rutherford2022normative}
Rutherford, S., Kia, S.~M., Wolfers, T., Fraza, C., Zabihi, M., Dinga, R.,
  Berthet, P., Worker, A., Verdi, S., Ruhe, H.~G., et~al. (2022).
\newblock The normative modeling framework for computational psychiatry.
\newblock {\em Nature {P}rotocols}, 17(7):1711--1734.

\bibitem[Staicu et~al., 2012]{staicu2012skewedFDA}
Staicu, A.-M., Crainiceanu, C.~M., Reich, D.~S., and Ruppert, D. (2012).
\newblock Modeling functional data with spatially heterogeneous shape
  characteristics.
\newblock {\em Biometrics}, 68(2):331--343.

\end{thebibliography}

\end{document}